\documentclass[preprint,prb,citeautoscript]{revtex4}

\usepackage{amssymb}
\usepackage{amsmath}
\usepackage{amsfonts}
\usepackage{graphicx}

\newcommand{\lbrs}{\left[}
\newcommand{\rbrs}{\right]}

\renewcommand{\t}[1]{\mbox{\boldmath$#1$}}
\renewcommand{\d}{{\text{d}}}

\newcommand{\D}[1]{\frac{\d~}{\d #1}}
\newcommand{\DD}[1]{\frac{\d_2~}{\d #1^2}}

\newcommand{\Tr}[1]{{{\frac{1}{N}\text{Tr}}\lbrs #1 \rbrs}}

\newcommand{\aeq}{\approx}

\newcommand{\vdW}{\text{vdW}}
\newcommand{\DFT}{\text{DFT}}
\newcommand{\damp}{d} % {\text{damp}}
\newcommand{\eV}{{\text{eV}}}
\newcommand{\meV}{{\text{meV}}}
\newcommand{\Ang}{\text{\AA}}
\newcommand{\etal}{\emph{et al}}

\newcommand{\eref}[1]{equation (\ref{#1})}

\newcommand{\erefs}[2]{equations \ref{#1}-\ref{#2}}
\newcommand{\rcite}[1]{ref.\ [\onlinecite{#1}]}

\newcommand{\chib}{{\bar{\chi}}}
\newcommand{\wb}{{\bar{w}}}
\newcommand{\cwb}{{\mathcal{C}}}

\newcommand{\FF}{\mathcal{F}}

\newcommand{\zp}{{z^{\prime}}}
\newcommand{\tC}{\tilde{C}}

\begin{document}
\title{A theoretical and semiemprical correction to the long-range
dispersion power law of stretched graphite}
\author{Tim Gould}
\email{t.gould@griffith.edu.au}
\author{Ken Simpkins}
\author{John F. Dobson}
\affiliation{Nanoscale Science and Technology Centre, Nathan campus,
Griffith University, 170 Kessels Road,  Nathan, QLD 4111, Australia}
\begin{abstract}
In recent years intercalated and pillared graphitic systems
have come under increasing scrutiny because of their potential
for modern energy technologies. While traditional \emph{ab initio}
methods such as the LDA give accurate geometries for graphite
they are poorer at predicting physicial properties such as cohesive
energies and elastic constants perpendicular to the layers
because of the strong dependence on long-range dispersion forces.
`Stretching' the layers via pillars or intercalation further
highlights these weaknesses. We use the ideas developed by
[J. F. Dobson \etal, Phys. Rev. Lett. {\bf 96}, 073201 (2006)]
as a starting point to show that the asymptotic
$C_3 D^{-3}$ dependence of the cohesive energy on layer spacing $D$ in
bigraphene is universal to all graphitic systems with evenly
spaced layers. At spacings appropriate to intercalates, this
differs from and begins to dominate the $C_4 D^{-4}$
power law for dispersion that has been widely used previously.
The corrected power law (and a calculated $C_3$ coefficient) is then
unsuccesfully employed in the semiempirical approach of
[M. Hasegawa and K. Nishidate, Phys. Rev. B {\bf 70}, 205431 (2004)] (HN).
A modified, physicially motivated semiempirical method including
some $C_4 D^{-4}$ effects allows the HN method to be used successfully
and gives an absolute increase of about $2-3\%$ to the predicted
cohesive energy, while still maintaining the correct
$C_3 D^{-3}$ asymptotics.
\end{abstract}

\maketitle

\section{Introduction}
The graphite form of carbon is a discretely layered material.
The $sp^2$ hybridised orbitals keep the layers in a rigid hexagonal
pattern while the $\pi_z$ orbitals help bind the layers. This
weak interlayer binding gives graphite a small elastic constant
($c_{33}$) perpendicular to the plane which allows graphite to be
`stretched' by pillaring (see eg. \rcite{Morishige2005}) and
intercalation (see eg. \rcite{IntercalateDefn1994})
by other substances with potentially useful
applications for Hydrogen storage and other new energy technology.

Standard density functional theory (DFT) \cite{DFT} based
approaches such as the LDA and GGA\cite{GGA}
are known (see \rcite{Hasegawa2004} for a summary)
to have problems predicting the interlayer binding energy and
interlayer elastic constant of graphite at its experimental
layer separation. This is presumed to be caused by the inability of
these functionals to accurately include the long-range London
dispersion forces (often denoted van der Waals forces in DFT papers,
a notation we adopt to maintain consistency with other work).
LDA/GGA correspondingly predict an exponentially decreasing binding
energy for $D\gg D_0$ (where $D$ is the
interlayer separation distance and $D_0=3.337\Ang$ is the experimental
interlayer separation distance) as opposed to the correct power
law behaviour.

Various authors \cite{Hasegawa2004,Girifalco2002,Rydberg2003,%
Dappe2006,Ortmann2006,Hasegawa2007,Rydberg2003} have proposed
corrections to the LDA/GGA results that yield an additional long-range
attractive layer-layer potential of the form $C_4 D^{-4}$. By contrast
Dobson, White and Rubio (DWR)\cite{Dobson2006} have shown that
the asymptotic power law behaviour for bigraphene is $C_3 D^{-3}$ due
to its unusual bandstructure near the $K$ point%
\cite{Wallace1947,SaitoDresselhausBook},
suggesting that even these \emph{ab initio} and semiempirical
corrections to LDA/GGA miss some important physics.

In this work we first show that the $C_3 D^{-3}$ power law is universal
to many-layered graphitic systems with uniform interlayer separation,
including those with an infinite number of layers such as rare gas
intercalated \footnote{It can be shown that in the case of an
insulating intercalate sitting \emph{on} each graphene layer that
the asymptotic form of the dispersion is $C_3 D^{-3}$ with $C_3$
unchanged from the non-intercalated but stretched bulk case. It seems
highly likely that this result would be the same for a non-interacting
insulating layer situated between the layers.} or pillared
graphite.

We then use our energy expression to calculate the correct
$C_3$ coefficient for bulk graphite and adapt the method of
Hasegawa and Nishidate (HN)\cite{Hasegawa2004} to emply
a corrected power law, thereby permitting empirical modelling of
the non-asymptotic region when $D\aeq D_0$. This investigation
suggests that the
different power-law and coefficient could have effects on
semiempirical and other methods which assume a $C_4 D^{-4}$ decay
of the dispersion potential but that such effect may dominate only
in for $D>D_0$.

\section{Asymptotic power law}

The success of the random-phase approximation (RPA)
in generating a correlation energy functional through the
Adiabatic Connection Formula and Fluctuation-Dissipation Theorem (ACFFDT)
with the correct power law for long-range dispersion forces
is well studied\cite{DobsonChap1994,Pitarke1998,
Dobson1999,Furche2001,Fuchs2002,Miyake2002,GouldThesis,Jung2004,Marini2006}.
For the case of graphene compounds, DWR\cite{Dobson2006}
used a long-wavelength approximation
to the bare density-density response ($\chi_0$) function of graphene
to prove a $C_3 D^{-3}$ dispersion potential for bigraphene while also
reproducing known results for other materials through
the same method.

If we assume (as in DWR) that the in-plane response of a graphene plane can
be approximated for low surface-parallel wavenumber ($q$)
by a homogenous system of similar physics then we can write the RPA
equation for the interacting density-density response ($\chi$) as
follows:
\begin{align}
&\chi_{\lambda}(q,z,\zp;u) = \chi_0(q,z,\zp;u)
\nonumber \\
&\hspace{5mm}
+ \lambda\int \d x \d y \chi_0(q,z,x;u) w(q,x,y) \chi_{\lambda}(q,y,\zp;u)
\label{eqn:RPA}
\end{align}
where the integrals are one-dimensional and $\lambda$ is a coupling
constant to be used in the adiabatic connection formula. In the case
of a layered system where each layer is highly localised in $z$ space and
separated by a distance $D$ so that
$\chi_0(q,z,z';u)=\sum_{i=0}^{N-1}\chib(q;u)\delta(z-z')\delta(z-iD)$
we may rewrite \eref{eqn:RPA} as a tensor equation over layer indices
$i$ and $j$
\begin{align}
\t{\chi}_{\lambda}(q,u;D) =& \chib(q,u)\t{1}
\nonumber\\
&+ \lambda \chib(q,u) \wb(q) \t{\Omega}(qD)\t{\chi}_{\lambda}(q,u;D)
\end{align}
where $\wb(q)=\frac{e^2}{2\epsilon_0 q}$ and $[\t{\Omega}]_{ij}=
\omega_{i-j}=e^{-qD|i-j|}$ ($0\leq i,j <N$) so that
$\chi_{\lambda}(q,z,z';u)=\sum_{ij}[\t{\chi}(q,u;D)]_{ij}\delta(z-iD)
\delta(z'-jD)$.

We can use the ACFFDT to write the correlation energy per layer
of a two-dimensionally homogeneous system as
\begin{align}
E_c=&-\frac{\hbar}{4\pi^2} \int_0^1 d\lambda \int_0^{\infty} \d u
\int_0^{\infty} q \d q
\int_{-\infty}^{\infty} \d z \int_{-\infty}^{\infty} \d z'
\nonumber \\
&
\times \lbrs \chi_{\lambda}(q,z,z';u) - \chib(q,z,z';u)\rbrs
\wb(q)e^{-q|z'-z|}.
\end{align}
Remembering that dispersion comes entirely from inter-layer correlation
effects and making use of the delta functions thus lets us calculate the
energy per unit area per layer of an $N$-layered system through
\begin{align}
U_{\vdW}=&-\frac{\hbar}{4\pi^2} \int_0^1 d\lambda \int_0^{\infty} \d u
\int_0^{\infty} q \d q
\nonumber\\
&\hspace{10mm}
\times \lbrs \FF_{\lambda}(q,u;D) - \FF_{\lambda}(q,u;\infty) \rbrs
\end{align}
where
\begin{align}
\FF_{\lambda}(q,u;D)=&\wb(q)\Tr{\t{\chi}_{\lambda}(q,u;D)\t{\Omega}(qD)}.
\end{align}

Due to the high level of symmetry $\t{\Omega}$ takes the form of a
Toeplitz matrix. This allows us to make use of Szeg\"o's Theorem
(\rcite{Toeplitz} contains a good review of Szeg\"o's Theorem and its
applications) to calculate the trace in the limit $N\to\infty$ (these
equations can also be obtained by Fourier methods).
Defining
\begin{align}
\tau(\xi)=\sum_{k=-\infty}^{\infty} \omega_{k} e^{i k \xi}
=\frac{\sinh(qD)}{\cosh(qD)-\cos(\xi)}
\end{align}
as the Fourier transform of the tensor elements of $\t{\Omega}$
we then find
\begin{align}
\FF_{\lambda}(q,u;D)&=\Tr{(\t{1}-\lambda\cwb\t{\Omega})^{-1}\cwb\t{\Omega}}
\nonumber\\
&=\frac{1}{2\pi}\int_{-\pi}^{\pi}\d\xi
\frac{\cwb\tau(\xi)}{1-\lambda \cwb\tau(\xi)}
\nonumber\\
&=\frac{\sinh(qD)\cwb}{\sqrt{[\cosh(qD)-\lambda \cwb\sinh(qD)]^2-1}}
\end{align}
where $\cwb=\chib(q,u)\wb(q)$.

For stretched graphitic systems the dominant energy contribution
of $\chi$ occurs
when $q$ and $u$ are small so that we can approximate the bare response
by its small $q$ and $u$ expansion
$\chib(q,u)\aeq -(2\hbar)^{-1} q^2(v_0^2 q^2 + u^2)^{-1/2}$
as calculated by DWR and Eqn. 3 of \rcite{Gonzalez2001}
We can now write $\cwb=-\kappa[1+u^2/(v_0 q)^2]^{-1/2}$ where
$\kappa=\frac{e^2}{4\epsilon_0\hbar v_0}=12.1$ for graphene where
$v_0=5.7\times 10^5\text{ms}^{-1}$.

If we make changes of variables $\theta=qD$ and $\sinh(\eta)=\frac{u}{v_0q}$
(so that $\cwb=-\kappa/\cosh(\eta)$) then we can eliminate $D$ from
inside the integrals\footnote{This change
of variables can also be made in the energy functional of any
finite number of equally spaced graphene layers making the
$D^{-3}$ power-law universal for evenly spaced systems}. We thus obtain
the energy expression
\begin{align}
U_{\vdW}=&\frac{\hbar v_0}{4\pi^2 D^3} \int_0^1 \d\lambda
\int_0^{\infty} \theta^2 d\theta
\int_0^{\infty} \cosh(\eta) \d\eta
\nonumber\\
&\hspace{15mm}\times
[\FF_{\lambda}(\theta,\eta) + \kappa(\cosh(\eta) + \lambda \kappa)^{-1}]
\label{eqn:Udisp}
\\
&=C_3 D^{-3}
\end{align}
with
\begin{align}
\FF_{\lambda}(\theta,\eta)=\frac{-\kappa \sinh(\theta)}
{\sqrt{[\cosh(\theta)\cosh(\eta) + \lambda \kappa \sinh(\theta)]^2
- \cosh(\eta)^2}}
\label{eqn:Fdisp}
\end{align}
and where the second term of \eref{eqn:Udisp} arises from letting
$D\to\infty$ in \eref{eqn:Fdisp}

Equation \ref{eqn:Udisp} is independent of $D$ aside from the desired
$D^{-3}$ term so that $C_3=D^3 U_{\vdW}$ depends only on $\kappa$.
For the graphitic case where $\kappa=12.1$ we find
\begin{align}
C_3&=2.12\times 10^{-2} \frac{e^2}{4\pi\epsilon_0}
= 0.80 \eV\Ang^3/\text{atom}.
\end{align}
By contrast the $C_4$ coefficient
predicted by Girifalco \etal \cite{Girifalco2002} is
$C_4 = 9.7949 \eV\Ang^4/\text{atom}$ which gives a potential
approximately four times ($0.079\eV$ vs $0.022\eV$) as large as that
of the inverse cubic power law at the experimental
interlayer spacing $D_0=3.337\Ang$
(equivalently this means that $C_3 D^{-3}>C_4 D^{-4}$
for $D > 4D_0$).

\section{Non-asymptotic behaviour}
\label{sec:Nonasymp}

While the $C_3 D^{-3}$ power law will certainly be the dominant
contributor to the dispersion potential for $D \gg D_0$,
the intermediate-range (when $D\aeq D_0$) will include a number of other
correlation effects. These include the $C_4 D^{-4}$ potential from
the atomic polarisibilities in the $z$ direction
in addition to a $C_{5/2}(D) D^{-5/2}$
potential from the metallic electrons promoted from the $\pi_z$ orbitals
due to layer overlap and hopping.
As $C_{5/2}(D)$ comes entirely from overlap of the $\pi_z$ orbitals
it ought to be derivable from an analysis of the band-structure. It is
expected to decay as an inverse exponential in $D$ due to localisation
of the $\pi_z$ orbitals. The $C_4$ coefficient should be
largely independent of $D$ although some electrons will be promoted
to metallic and graphitic response.

With such a varied collection of correlation effects it seems unlikely
that any simple \emph{ab initio} method will adequately include the
physics in the intermediate-range. Full RPA-ACFFDT calculations would
be expected to provide a seamless potential through a wide-range of $D$
however these are extremely difficult with current numerical approaches:
for example, the van der Waals energetics of the semiconducting
layered boron nitride system have been described succesfully using
RPA energies\cite{Marini2006}, but graphite gives convergence
difficulties\cite{RPAGraphiteProblems}.

\section{Semiempirical method}

LDA calculations are expected to yield fairly accurate total energies for
graphene when the interlayer spacing is compressed from its equilibrium
value. Likewise the $C_3 D^{-3}$ dispersion potential is expected to be
accurate for layer spacing much greater than that of equilibrium. The
intermediate range is more difficult to predict with neither method
dealing sufficiently with the physics in that region.

The method proposed in HN\cite{Hasegawa2004} gives a fairly simple
means (with minimal empirical contribution)
of connecting the two regimes through the use of a fitting
function. It is a semiempirical approach as the fitting function has
its parameters chosen by matching experimental values for the lattice
spacing and elastic constant $c_{33}$. While this method predicts a
reasonable value for the cohesion energy of graphite it, as with other
methods, does not exhibit the correct behaviour in the tail due to the
incorrect use of a $C_4 D^{-4}$ type dispersion law.
In order to maintain consistency with this earlier work we re-examine
the major results of their paper utilising the correct
$C_3 D^{-3}$ dispersion law. To further maintain consistency we use
the parametrisation of the LDA and GGA from the same paper.

\subsection{Semiempirical approach with pure {$C_3 D^{-3}$} dispersion}
For our first new approach we adapt Equation 5 of HN to include
the corrected form of the dispersion potential
\begin{align}
U(D)=&\lbrs 1 - f_{\damp}(D) \rbrs U_{\DFT}(D)
%\nonumber\\
%&\hspace{10mm}
+ f_{\damp}(D) U_{\vdW}(D)
\label{eqn:FitFn1}
\end{align}
where $U_{\vdW}(D)=C_3 D^{-3}$. Following HN, we use a Thomas-Fermi
damping function 
\begin{align}
f_{\damp}(D)=[1 + e^{-(D-D_W)/\delta}]^{-1}
\label{eqn:FitFn2}
\end{align}
where $D_W$ and $\delta$ are free parameters. The term involving
$\Delta\zeta(4)$ is absent due to our $C_3$ coefficient being sourced
from a bulk rather than a sum over pairwise potentials for
multiple layers.
$U^{\DFT}(D)$ is the parametrised LDA or GGA potential taken
from equation 2 of HN.

As in HN we attempted to determine $\delta$ and $D_W$ by ensuring that
$\D{D}U(D_0)=0$ and $\DD{D}U(D_0)=c_{33}/(\rho D_0)$ where
$c_{33}=40.7$GPa, $D_0=3.337\Ang$ and $\rho=0.382\Ang{}^{-2}$ take their
experimental values (from \rcite{Gauster1974} for $c_{33}$ and
\rcite{Baskin1955} for $\rho$ and $D_0$). Using
$U_{\vdW}=0.80 \meV \Ang^3 D^{-3}$ we find that the HN fitting equations do
not have a solution for the
LDA or GGA. This lack of solution is not unexpected as the
lack of other dispersion terms is expected to
underestimate the dispersion for values of $D\aeq D_0$.

\subsection{Semiempirical approach with mixed {$C_3 D^{-3}$} and
{$\tC_4 D^{-4}$} dispersion}
While the $C_3 D^{-3}$ term will certainly dominate over $C_4 D^{-4}$
for $D\gg D_0$ we know that it insufficiently models the physics
for $D \aeq D_0$, which we believe to be the cause of the fitting
problems with the HN method for the semi-empirical method given in (A)
above. The $C_4=9.795\meV\Ang^4$ coefficient used in HN is
derived from a $C_6=16.34\meV\Ang^6$ coefficient calculated by Girifalco
\etal\cite{Girifalco2002} and constructed to ensure good Lennard-Jones
modelling for a wide variety of graphitic systems. As such we
propose to use its presumed accuracy for $D\aeq D_0$ as a correction
to our $C_3 D^{-3}$ van der Waals function in order to better
include the intermediate range physics.

The simplest way to do this is to assume a correction to our function
of the form $\tC_4 D^{-4}$.
The $\tC_4$ term is introduced firstly to cover the dispersion
interaction due to the polarizability of the $\pi_z$ and $sp^2$ electrons
in the $z$ direction perpendicular to the graphene planes, and
polarisability of the $sp_2$ electrons parallel to the plane. These
contributions to the dispersion interaction do not require
long-wavelength collective electronic motions and
therefore\cite{Dobson2006} are presumably describable by conventional
asymptotics.  These interaction are not included in our $C_3 D^{-3}$ term,
which is solely due to polarizability of the $\pi_z$ electrons along
the graphene planes. The $\tC_4$ term also has to account for the doped,
metallic nature of the graphene planes near $D\aeq D_0$ due to overlapping
of electron bands arising from the hopping of electrons from layer to layer.
The corresponding attraction depends on the doping level, which decays
exponentially with D.  While this is not a $D^{-4}$ law, it does decay
faster than $D^{-3}$ and hence is reasonably represented.

Accordingly we now choose $\tC_4$ so that the total van der Waals potential
at the experimental lattice spacing remains the same in the two methods.
This implies that
\begin{align}
C_4 D_0^{-4} = C_3 D_0^{-3} + \tC_4 D_0^{-4}
\end{align}
which is true for $\tC_4=7.12\meV\Ang^4$. This correction ensures we
maintain similar $D\aeq D_0$ behaviour while obtaining a correct power
law for $D\gg D_0$. The van der Waals potential now takes the form
\begin{align}
U_{\vdW}(D)=C_3 D^{-3} + \tC_4 D^{-4}.
\end{align}

\begin{figure}[htb]
\includegraphics[width=8.5cm]{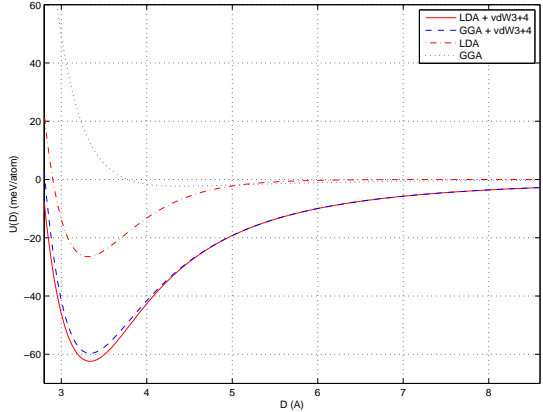}
\caption{Potential energy versus lattice spacing ($D$). The solid line
is the LDA corrected by the $C_3 D^{-3} + \tC_4 D^{-4}$ while the
dashed line is the
corrected GGA. The dash-dot and dotted lines are the pure LDA and GGA
respectively.}
\label{fig:VarStarts34}
\end{figure}
In order to ensure that \erefs{eqn:FitFn1}{eqn:FitFn2} correctly
match the empirical data we must set
$\delta=0.221$, $D_W=3.283$ for the LDA and
$\delta=0.340$, $D_W=3.019$ for the GGA when using the HN
fitting function. Figure \ref{fig:VarStarts34} shows the effect of this
combined fit on both the LDA and GGA.

\begin{figure}[htb]
\includegraphics[width=8.5cm]{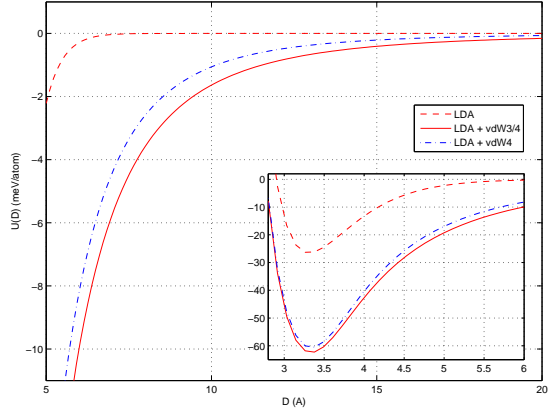}
\caption{Potential energy versus lattice spacing ($D$). The dashed line
is the uncorrected LDA, the solid line
is the LDA corrected by $C_3 D^{-3} + \tC_4 D^{-4}$
while the dash-dots are the LDA corrected
by $C_3 D^{-4}$ as in HN. The inset shows the behaviour near $D_0$ while
the main graph shows the different asymptotics.}
\label{fig:vdWThreeTypes}
\end{figure}
In Figure \ref{fig:vdWThreeTypes} we show, for the LDA case,
 a more detailed comparison of three methods
(the LDA, that of HN and the second method proposed here). It is quite clear
from the graph that the method proposed here with the extra
$\tC_4 D^{-4}$ correction closely matches that of HN for $D\aeq D_0$ but
maintains different asymptotes for $D\gg D_0$. This suggests that
the $C_3 D^{-3} + \tC_4 D^{-4}$ approximation, while a somewhat
crude model of the true physics in the electron density overlap
region, is able to maintain consistency with other methods.

The most `measurable' effect of the semiempirical approach is the
interlayer cohesive energy $\min(U(D))$. Table \ref{tab:Ecoh} summarises
the results from HN with the addition of the new results calculated here.
The new power law, used as the sole dispersion term does not give
a valid cohesive energy due to the lack of a solution to the fitting function
for both the LDA and GGA. The effect of the $\tC_4 D^{-4}$ correction
to the $C_3 D^{-3}$ van der Waals functional on the cohesive energy
is to give a very close cohesive energy to those predicted by HN,
differing by only $1.7\eV$ for the LDA and $2.3\eV$ for the GGA or about
$2-3\%$.
\begin{table}[hbt]
\begin{tabular}{p{8mm}p{21mm}p{21mm}p{21mm}}
\hline
& LDA/GGA & Expt${}^a$ & Expt${}^b$
\\
$U_{coh}$ & 26.5/2.3  & $52.5\pm 5$ & $35^{+15}_{-10}$
\\
\hline
& L/G-vdW3 & L/G-vdW4 & L/G-vdW3+4
\\
$U_{coh}$ & -/- & 60.7/57.4 & 62.4/59.7
\\
\hline
\end{tabular}
\caption{Cohesive energies per atom calculated by various approximations
(in $-\meV$). L/G-vdW3 is the LDA/GGA with a $C_3 D^{-3}$
correction (for which meaningful results do not exist)
while L/G-vdW4 has a $C_4 D^{-4}$ correction
(taken from HN using $C_6=16.34\eV\Ang^6$) and L/G-vdW3+4 has the
combined correction $C_3 D^{-3} + \tC_4 D^{-4}$.
The experimental results are taken from \rcite{Zacharia2004}
for \emph{a} and \rcite{Benedict1998} for \emph{b}.}
\label{tab:Ecoh}
\end{table}

\section{Conclusions and further work}

In this paper we have investigated the asymptotic dispersion potential
of `stretched' graphite and found it to obey a $C_3 D^{-3}$ type power
law as opposed to the commonly employed $C_4 D^{-4}$. This places it
in the same class of power law as bigraphene but in a different class
to layered insulators ($D^{-4}$ power laws) and layered metals ($D^{-5/2}$).
This result has important implications for semiempirical (and otherwise)
corrections to the LDA/GGA which have employed an incorrect power law.

Furthermore we have employed the corrected power law in the simple
semiempirical method of Hasegawa and Nishidate\cite{Hasegawa2004}
and found that, used as the sole dispersion term,
it will not allow a valid solution to the HN fitting function.
Reinclusion of a reduced $\tC_4 D^{-4}$ term to include other physics from
the non-asymptotic regime allows the method to be employed
and gives similar results to HN for $D\aeq D_0$
whilst ensuring the correct asymptotic behaviour is maintained. Its
effect on the cohesive energy is fairly minimal with an absolute
increase of the predicted cohesive energy of approximately $2-3\%$.

While we believe that this power law (and the semi-empirical correction
to it) should be accurate and useful for
large layer spacings as in non-metallic intercalates and pillared systems
we are not convinced that it will be as accurate
in predicting the behaviour in the intermediate range of spacings
without correction for other effects. Accurate RPA-ACFFDT calculations
would provide a valuable benchmark for this and other methods.
Until such time as these are available we hope that semi-emprical
techniques like that discussed here should improve the accuracy of
LDA and GGA calculations with widely spaced graphene layers.

\section{Acknowledgements}
The authors would like to thank Evan Gray for fruitful
discussions. This research was conducted under a Discovery Grant for
the Australian Research Council.

\bibliographystyle{apsrev}
\bibliography{vanDerWaals,DFT,Wannier,Misc,Experiment,Graphene}

\end{document}